\documentclass{aip-cp}

\usepackage[numbers]{natbib}
\usepackage{rotating}
\usepackage{graphicx}

\begin{document}
\def\a{{\alpha }}
\def\g{{\gamma }}
\def\b{{\beta }}
\def\z{{\zeta }}
\def\zab{{\zeta_{ab}}}
\def\be{\begin{equation}}
\def\ee#1{\label{#1}\end{equation}}
\def\d{\textsf{d} }
\def\c{\textsf{c} }
\def\e{\textsf{e} }
\def\be{\textsf{b} }
\def\s{\textsf{s} }
\def\x{\textsf{x} }
 \def\bx{\mathbf{x} }
 \def\bp{\mathbf{p} }
 \def\p{\textsf{p} }
 \def\I{\textsf{I} }
  \def\n{\textsf{n} }
 \def\pp{\textsf{P} }
 \def\q{\textsf{q} }
\def\k{\textsf{k} }
 \def\J{\textsf{J} }
\def\no{\nonumber}
\def\lb{\label}
\def\h{\textsf{h} }
\def\x{\textsf{x} }
\def\D{\textsf{D}}
\def\lb{\label}
\def\Ki{{\rm Ki}}
\def\lab{{\langle\alpha\beta\rangle}}
\newcommand{\ben}{\begin{eqnarray}}
\newcommand{\een}{\end{eqnarray}}

\title{Thermal Conductivity, Shear and Bulk Viscosities for a Relativistic Binary Mixture}

\author[aff1]{Valdemar Moratto}
\eaddress{moratto.valdemar@gmail.com}
\author[aff1]{Gilberto M. Kremer}
\eaddress{kremer@fisica.ufpr.br}

\affil[aff1]{Departamento de F\'{\i}sica, Universidade Federal do Paran\'a, Caixa Postal 19044, 81531-980 Curitiba, Brazil}

\maketitle

\begin{abstract}
In the present work, we deal with a binary mixture of diluted relativistic gases within the framework of the kinetic theory. The analysis is made within the framework of the Boltzmann equation. We assume that the gas is under the influence of an isotropic Schwarzschild metric and  is composed of particles with speeds comparable with the light speed. Taking into account the constitutive equations for the laws of Fourier and Navier-Stokes, we obtain expressions for the thermal conductivity, the shear, and bulk viscosities. To evaluate the integrals we assume  a hard-sphere interaction along with non-disparate masses for the particles of each component. We show the analytical expressions and the behavior of the transport coefficients with respect to a relativistic parameter which gives the ratio of the rest energy of the particles to the thermal energy of the gas. We also determine the dependence of the transport coefficients with respect to the gravitational potential and demonstrate that the corresponding one component limit is recovered by considering particles with equal masses, in accordance with the kinetic theory of a single fluid.
\end{abstract}

\section{INTRODUCTION}
One of the most important outcomes of the kinetic theory is to provide analytic expressions for the transport coefficients for diluted gases. Appealing to astrophysical scenarios where the aforesaid theory can be applied, it may concern to the interstellar cloud gases and particular conditions in the interior of stars. In those cases, diluted gases can lie under the presence of a gravitational potential and within relativistic (high temperatures) conditions.

The study of the kinetic theory within the tenets of general relativity, in particular the study of constitutive equations and the corresponding transport coefficients in the presence of curved space-time is a subject that has not been studied deeply in the literature. We can mention some works in which some metrics were adopted to analyze a relativistic gas \cite{Ch1,Ch2,Bern}. Recently, the interest of the study of gases in the presence of gravity has raised \cite{And1,And2} and curved space-time have been incorporated to the Boltzmann equation to obtain constitutive equations \cite{KS,KD,MK}. Furthermore, it has been shown that the so-called Tolman law \cite{To1,To2} can be derived from the kinetic theory formalism \cite{AL} and that the transport coefficients depend on the gravitational potential gradient \cite{KS}. On the other hand, from the phenomenological point of view, the first post Newtonian approximation (1PN)  implies a correction to the pressure that depends on  the gravitational potential \cite{Wein}. Such a dependence has been obtained from a microscopic point of view by determining  the first post Newtonian approximation to the Maxwell-J\"uttner distribution function  \cite{KRW}. This last statement is also demonstrated in this article.

In the present work we start the analysis with the covariant Boltzmann equation \cite{CK} for a linear regime. The method to solve the Boltzmann equation is one that  combines the Chapman-Enskog \cite{CC,GMK} and Grad \cite{Grad} formalisms \cite{CEG,CK}, it leads to the determination of the constitutive equations for the linear fluxes of heat, particles and momentum. We have developed such a program in \cite{MK} and in this work we evaluate the thermal conductivity, bulk and shear viscosities. We show the behavior of such quantities with respect to the ratio of the rest energy of the particles and the thermal energy of the gas, the corresponding one-component and non-relativistic limits and ultimately their dependence with the gravitational potential.

\section{BASIC EQUATIONS AND METHOD OF SOLUTION}

Let us consider a binary mixture of ideal and relativistic diluted gases under the influence of a curved space-time described with the isotropic Schwarzschild metric:
\ben\label{1}
 ds^2=g_0(r)\left(dx^0\right)^2-g_1(r)\delta_{ij}dx^idx^j,\qquad
g_0(r)=\frac{\left(1-\frac{GM}{2c^2r}\right)^2}{\left(1+\frac{GM}{2c^2r}\right)^2},\qquad
g_1(r)=\left(1+\frac{GM}{2c^2r}\right)^4,
\een
here $G$ is the gravitational constant, $M$ the total mass of the spherical source and $r$ the radius.

We consider that the gas is constituted by particles that do not have internal degrees of freedom. Each
of these particles of the constituent $a=\{1,2\}$ have rest mass $m_{a}$
and are characterized by the space-time coordinates $x^{\mu}=\left(ct,\bf x\right)$
and  momentum $p_{a}^{\mu}=\left(p_{a}^{0},\textbf{p}_{a}\right)$. The mass-shell condition $g_{\mu\nu}p_{a}^{\mu}p_{a}^{\nu}=m_{a}^{2}c^{2}$
imposes the following restrictions for the contravariant and covariant components $p_{a}^{0}=p_{a0}/{g_{0}}$ and $p_{a0}=\sqrt{g_{0}\left(m_{a}^{2}c^{2} -g_{1} \vert{\bf p}_{a}\vert^{2}\right)}$, respectively.
The state of the gas is described by the one-particle distribution function $f_{a}\left(x^{\mu},\textbf{p}_{a}\right)$. This function has a statistical meaning because the quantity $f_{a}\left(x^{\mu},\textbf{p}_{a}\right){d}^{3}x\,{d}^{3}p_{a}$ is the number of particles of the constituent $a$ in the volume element between $\bf x$, ${\bf x}+{d}^{3}x$ and
$\textbf{p}_{a}$, $\textbf{p}_{a}+{d}^{3}p_{a}$ at the time $t$. The evolution of the distribution function is governed by the Boltzmann equation \cite{CK},
\ben\label{2}
p_{a}^{\mu}\frac{\partial f_{a}}{\partial x^{\mu}}-\Gamma_{\mu\nu}^{i}p_{a}^{\mu}p_{a}^{\nu}\frac{\partial f_{a}}{\partial p_{a}^{i}} =\sum_{b=1}^{2}\int(f_{a}'f_{b}'-f_{a}f_{b})F_{ba}\sigma_{ab}\, d\Omega\sqrt{-g}\frac{d^{3}p_{b}}{p_{b0}}.
\een
Here the metric connection is given through the Christoffel symbols $\Gamma_{\mu\nu}^{i}$ and usual quantities in the right hand side are to be defined. We have the invariant flux $F_{ba}=\sqrt{(p_{a}^{\mu}p_{b\mu})^{2}-m_{a}^{2}m_{b}^{2}c^{4}}$ and the invariant differential elastic cross-section $\sigma_{ab}d\Omega$ for collisions of species $a$ and $b$ where $d\Omega$ is the corresponding solid angle element. We have the invariant differential element $\sqrt{-g}\frac{d^{3}p_{b}}{p_{b0}}$ with $\sqrt{-g}=\hbox{det}\left[g^{\mu\nu}\right]$. Quantities denoted with a prime are evaluated with the momentum of the particles after a binary collision, that is, $f'_{a}\equiv f({\bf x},{\bf p}'_{a},t)$.

It is well-know that the solution of the Boltzmann equation in a situation in which the collisions do not alter the distribution function is given by the Maxwell-J\"uttner distribution function \cite{JUTTNER}. That is, the distribution function that describes the local equilibrium reads

\begin{equation}\label{3}
f_{a}^{(0)}=\frac{\n_{a}}{4\pi kTm_{a}^{2}c{K}_{2}\left(\zeta_{a}\right)} \exp\left(-\frac{U_{\mu}p_{a}^{\mu}}{kT}\right)=\frac{\n_{a}}{4\pi kTm_{a}^{2}c{K}_{2}\left(\zeta_{a}\right)} \exp\left(-\frac{c\sqrt{m_{a}^{2}c^{2}+g_{1}\vert{\bf p}_{a}\vert^{2}}}{kT}\right).
\end{equation}
The second equality of Eq. (\ref{3}) is evaluated in the co-moving frame, where $U^{\mu}=\left(c/\sqrt{g_{0}},\textbf{0}\right)$. Furthermore, $K_2(\zeta_a)$ is a modified Bessel function of second kind and $\zeta_a=m_ac^2/kT$.

Following the standard procedures \cite{CK} of the kinetic theory, the obtention of the balance equations results from the successive multiplication of the Boltzmann equation with the moments of the distribution function and integration over $\sqrt{-g}\frac{d^{3}p_{a}}{p_{a0}}$. Such development is an unnecessary task to be done here due to the objectives of the present work. Here we recall the definition of the energy-momentum tensor
\ben\label{4}
T_a^{\mu\nu}=c\int p_{a}^{\mu}p_a^\nu f_{a}\sqrt{-g}\frac{d^{3}p_{a}}{p_{a0}},\qquad\hbox{for the mixture we have}\qquad T^{\mu\nu}=\sum_{a=1}^{2}T_a^{\mu\nu},
\een
and the particle four-flow of species $a$
\begin{equation}\label{5}
N_{a}^{\mu}=c\int p_{a}^{\mu}f_{a}\sqrt{-g}\frac{d^{3}p_{a}}{p_{a0}},\qquad\hbox{and for the mixture}\qquad N^{\mu}=\sum_{a=1}^{2}N_a^{\mu}.
\end{equation}
Following the Eckart frame \cite{Eck}, we introduce the following decomposition of
$N_{a}^{\mu}$ in terms of the hydrodynamic four-velocity $U^\mu$ as
\begin{equation}
N_{a}^{\mu}=\n_{a}U^{\mu}+\J_{a}^{\mu},\qquad\hbox{where}\qquad \n_{a}=\frac{N_{a}^{\mu}U_{\mu}}{c^{2}}
\label{6}
\end{equation}
is the local number of particles of species.
Above also appears the diffusive particle four-flux $\J_{a}^{\mu}$ defined as
\begin{equation}\label{6.4}
\J_{a}^{\mu}=\Delta_{\nu}^{\mu}c\int p_{a}^{\nu}f_{a}\frac{d^{3}p_{a}}{p_{a0}},\qquad \hbox{with the projector} \qquad \Delta^{\mu\nu}=g^{\mu\nu}-\frac{1}{c^{2}}U^{\mu}U^{\nu},\qquad \hbox{and it holds}\qquad \J_{a}^{\mu}U_{\mu}=0.
\end{equation}

As usual, the energy-momentum tensor Eq. (\ref{4}) can be decomposed in an irreducible form as
\ben
T_a^{\mu\nu}=\frac{\n_a\e_a}{c^{2}}U^{\mu}U^{\nu}
+\frac{1}{c^{2}}U^{\mu}\left(\q_a^{\nu}+\h_a\J_{a}^{\nu}
\right)+\frac{1}{c^{2}}U^{\nu}\left(\q_a^{\mu}+\h_a\J_{a}^{\mu}
\right)-(\p_a+\varpi_a)
\Delta^{\mu\nu}+\p_a^{\langle\mu\nu\rangle},\label{6.5}
\een
where $\e_a$ is the energy per particle, $\p_a$ the pressure and $\h_a=\e_{a}+\frac{\p_{a}}{\n_{a}}$  the enthalpy per particle of constituent $a$.

From equation (\ref{6.5}), with some appropriate projections, we are able to obtain general expressions for the partial heat four-flux $\q^{\mu}_a$, dynamical pressure $\varpi_a$ and pressure deviator tensor $\p_a^{\langle\mu\nu\rangle}$ as:
\ben\label{8}
\q_a^{\mu}+\h_a\J_{a}^{\mu}=\Delta_\sigma^\mu T_{a}^{\sigma\nu}U_{\nu},\qquad
\p_{a}+\varpi_a=-\frac13\Delta_{\mu\nu}T_{a}^{\mu\nu}\qquad\hbox{and}\qquad
\p_a^{\langle\mu\nu\rangle}=\left(\Delta_{\sigma}^{\mu}\Delta_{\tau}^{\nu}
-\frac{1}{3}\Delta^{\mu\nu}\Delta_{\sigma\tau}\right)T_{a}^{\sigma\tau}.
\een

The method of solution of the relativistic covariant Boltzmann equation that we have used is a combination that mixes features of the Chapman-Enskog \cite{CC} and Grad \cite{Grad} formalisms (see \cite{CEG,K1}). It consists essentially in doing an expansion to first order of the distribution function that is solution of Boltzmann's equation for each species. Then an imposition of its compatibility with the solution given by Grad to linear regime leads to a linearization of Boltzmann equation for the thermodynamic four-fluxes of heat, particles, dynamic pressure and pressure deviator tensor. Here it is not viable to rewrite such an expression because it is quite long, but the reader  can find the complete analysis in \cite{MK}.

\section{FOURIER LAW}

Once upon the method described in last section is developed to linearize the Boltzmann equation, it leads to an algebraic system of equations for the thermodynamic fluxes. By taking the sum of the partial heat flux Eq. (\ref{8}a) we can obtain the Fourier law, which establishes that the thermal conductivity is the ratio between the total heat flux and the thermal force, in this case we obtain
\ben\label{9}
\q^\mu=\lambda\nabla^\mu\mathcal{T},\qquad \hbox{where}\qquad
\lambda=-\frac{\mathcal{H}_{11}+\mathcal{H}_{22}-2\mathcal{H}_{12}}
{T(\mathcal{H}_{11}\mathcal{H}_{22}-\mathcal{H}_{12}^2)}
\een
is the thermal conductivity. The generalized thermal  force has been defined as
\ben\label{10}
\nabla^\mu\mathcal{T}=\nabla^\mu T-\frac{T}{c^{2}}\Delta^{\mu i}\left[U^{\nu}\frac{\partial U_{i}}{\partial x^{\nu}}-\frac{1}{1-{\Phi^2}/4c^{4}}\frac{\partial{\Phi}}{\partial x^{i}}\right].
\een
This thermodynamic force has the following dependence:
\begin{itemize}
\item The first term is the gradient of the temperature as a legitime thermodynamic variable.
\item The second term has a contribution of the hydrodynamic four-acceleration in accordance with the phenomenological work developed by Eckart, see \cite{Eck}. This is a strictly relativistic term due to the factor $\sim T/c^2$. It represents an isothermal heat flux when matter is under acceleration and acts in opposite direction to the movement of the gas.
\item Lastly we observe a contribution to the heat flux due to the gradient of the gravitational potential $\Phi=-\frac{GM}{r}$. This is also a relativistic contribution $\sim T/c^2$.
\end{itemize}

An issue that deserves to be underlined is that equation (\ref{9}a) recovers the so-call Tolman law in the absence of heat flux and acceleration. That is, in the nearby of a gravitational source, a state of equilibrium of a relativistic gas can be achieved when the temperature gradient is counterbalanced by a gravitational potential gradient. If we evaluate the projector (Eq. (\ref{6.4}b)) with a Schwarzschild metric (Eq. (\ref{1})) in the comoving frame $U^{\mu}=\left(c/\sqrt{g_{0}},\textbf{0}\right)$ we have
\ben\label{11}
\Delta^{00}=0,\qquad \Delta^{ij}=-\frac{\delta^{ij}}{\left(1-\frac{\Phi}{2c^2}\right)^4}.
\een
Now, if we suppose a week field $\Phi/c^2\ll1$ we can expand the term
\ben\label{12}
\Delta^{ij}\simeq-\delta^{ij}\left[ 1+\frac{2\Phi}{c^2}+\frac{5\Phi^2}{2c^4} +\mathcal{O}\left(\frac{\Phi^3}{c^6}\right)\right],
\een
so that, from equation (\ref{10}) to lowest order we recover
\ben\label{ToL}
\frac{\nabla T}{T}=\frac{\nabla\Phi}{c^2}
\een
which is the Tolman law.

The $\mathcal{H}'s$ functions that appear in Eq. (\ref{9}b) are described as general functions in \cite{MK} and have been evaluated in \cite{KM} with the following two hypotheses:

\begin{enumerate}
\item The masses of the different constituents are similar, that is $m_2=m_1(1+\epsilon)$ where $\epsilon\ll1$.
\item We consider hard-spheres, and the diameters of the particles are constant and a small difference is assumed for the diameter of the species 2 with respect to the diameter of species 1 as: $\d_2=\d_1(1+\xi)$ with $\xi\ll1$. The hard-sphere differential cross-sections are given as functions of the diameters according to $\sigma_{11}=\d_1^2/4$, $\sigma_{22}=\d_2^2/4$ and $\sigma_{12}=(\d_1+\d_2)^2/16$. Furthermore, we obtain the relations: $\sigma_{11}=\sigma$, $\sigma_{12}=\sigma(1+\xi)$ and $\sigma_{22}=\sigma(1+2\xi)$ with $\sigma=$cte.
\end{enumerate}

In figure 1 we show the behavior of the dimensionless thermal conductivity coefficients for different situations of concentration as a function of the relativistic parameter $\z_1=m_1 c^2/kT$. We note that the thermal conductivity coefficient assume large values in the non-relativistic limit $\zeta_1\gg1$ than in the ultra-relativistic limit $\zeta_1\ll1$ and that it decreases by increasing the concentration ratio $n_2/n_1$.

\begin{figure}[h]
  \centerline{\includegraphics[width=270pt]{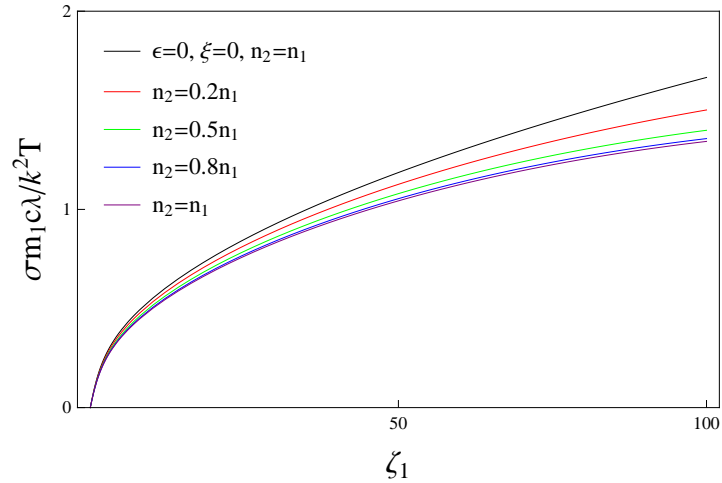}}
  \caption{(color on line) Thermal conductivity (dimensionless) as a function of  $\z_1=m_1 c^2/kT$. Black line represents the single fluid whereas colored curves are plotted with $\epsilon=0.01$ and $\xi=0.1$ and different concentrations.}
\end{figure}

It is worth to mention that we can recover the appropriate expression for the thermal conductivity in the case of a single fluid when we assume $m_1=m_2=m, \n_1=\n_2=\n, \sigma_{11}=\sigma_{22}=\sigma_{12}=\sigma$. In such a case, equation (\ref{9}b) reduces to \cite{CK,GLW}
\ben\label{14}
\lambda=\frac3{64\pi}\frac{ck}\sigma\frac{(\z+5G-G^2\z^2)^2\z^4K_2(\z)^2}{(\z^2+2)K_2(2\z)+5\z K_3(2\z)},
\een
here $G=K_3(\z)/K_2(\z)$  is an abbreviation for the ratio of modified Bessel functions of second kind (see \cite{CK}). The non-relativistic ($\z\gg$, low temperature) and ultra relativistic ($\z\ll1$, high temperature) limiting cases lead to
\ben\label{15}
\lambda=\frac{75}{64\d^2}\frac{k}{m}\sqrt{\frac{mkT}\pi} \left(1+\frac{13}{16\z}+\dots\right), \qquad \z\gg1,
\qquad 
\lambda=\frac{2ck}{\pi\d^2}\left(1-\frac{\z^2}{4}+\dots\right),\qquad \z\ll1.
\een
Here $\d$ represents the diameter of one particle.

\section{NAVIER-STOKES LAW}

In this section we show the constitutive equations for the transfer of momentum in the gas, these equations are commonly known as the Navier-Stokes law and they are obtained through the definitions written in Eqs. (\ref{8}b) and (\ref{8}c):
\ben\label{17}
\varpi=\sum_{a=1}^{2}\varpi_a=-\eta\nabla_\mu U^\mu,\qquad\qquad\hbox{and}\qquad\qquad \p^{\langle\mu\nu\rangle}=\sum_{a=1}^{2}\p_a^{\langle\mu\nu\rangle}=2\mu \nabla^{\langle\mu}U^{\nu\rangle}.
\een

First we analyze the bulk viscosity $\eta$, hence by following the methodology described in \cite{KM} we obtain
\ben\label{18}
\eta=\frac{kT\left(\mathcal{R}_{22}-\mathcal{R}_{21}\right)\p_1}
{c^3(\mathcal{R}_{11}\mathcal{R}_{22}-\mathcal{R}_{12}\mathcal{R}_{21})}
\left[\frac{\partial\ln \c_v^1}{\partial\ln\z_1}+\frac{\left(\mathcal{R}_{11}-\mathcal{R}_{12}\right)\n_2}{\left(\mathcal{R}_{22} -\mathcal{R}_{21}\right)\n_1}\frac{\partial\ln \c_v^2}{\partial\ln\z_2}\right].
\een
Here  the elements of the matrices $\mathcal{R}_{ab}$ are given in \cite{KM}. We have also introduced the derivative of the heat capacity per particle at constant volume $\c_v^a=k(\z_a^2+5G_a\z_a-G_a^2\z_a^2-1)$ as
\ben\label{19}
\frac{\partial\ln \c_v^a}{\partial\ln\z_a}=\frac{\z_a(20G_a+3\z_a
-13G_a^2\z_a-2G_a\z_a^2+2G_a^3\z_a^2)}{(1+G_a^2\z_a^2-\z_a^2-5G_a\z_a)},
\een
being $G_a=K_3(\z_a)/K_2(\z_a)$.

In figure 2, we show the plot of the bulk viscosity when the mass of the particles of species 2 are $m_2=m_1(1+\epsilon)$ with $\epsilon\ll1$ and the difference in the size of the molecules are $\d_2=\d_1(1+\xi)$, $\xi\ll1$, as in the last section.

\begin{figure}[h]
  \centerline{\includegraphics[width=270pt]{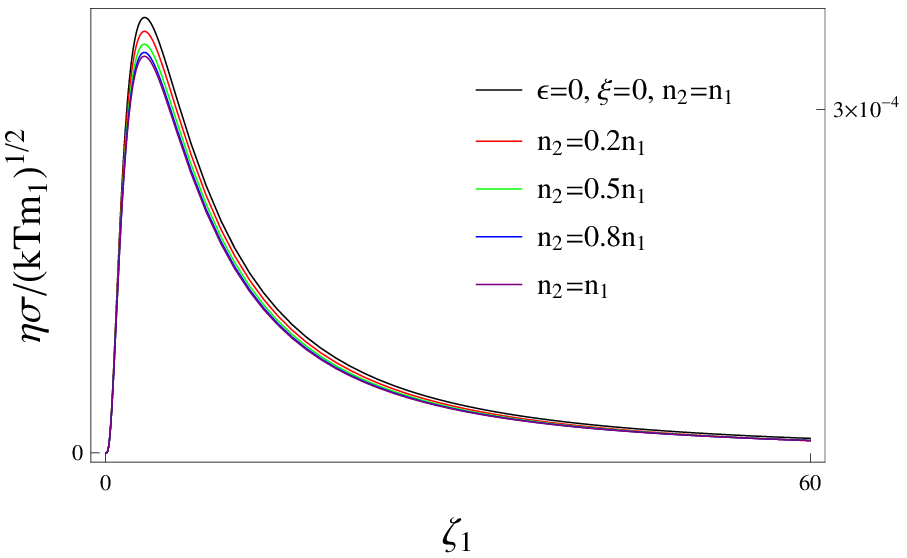}}
  \caption{(color on line) Bulk viscosity (dimensionless) as a function of  $\z_1=m_1 c^2/kT$. Black line represents the single fluid whereas colored curves are plotted with $\epsilon=0.01$ and $\xi=0.1$ and different concentrations.}
\end{figure}
From equation (\ref{18}) we can evaluate the one-component limit ($m_1=m_2=m, \n_1=\n_2=\n, \sigma_{11}=\sigma_{22}=\sigma_{12}=\sigma$) which yields
\ben\label{20}
\eta=\frac1{64\pi}\frac{kT}{c\sigma}\frac{\z^4K_2(\z)^2 (20G+3\z-13G^2\z-2G\z^2+2G^3\z^2)^2}{(2K_2(2\z)+\z K_3(2\z))(1-5G\z-\z^2+G^2\z^2)^2},
\een
according with the known expression \cite{CK,GLW}.
We can conclude from Fig. 2 that the bulk viscosity decreases by increasing the concentration of constituent labeled by 2 with respect to the one labeled by 1 and that the corresponding non and ultra-relativistic values tend to zero. Indeed, by evaluating the limiting non and ultra-relativistic cases we obtain
\ben\label{21}
\eta=\frac{25}{64\d^2\z^2}\sqrt{\frac{mkT}\pi}\left(1- \frac{183}{16}\frac{1}{\z^2}+\cdots\right), \quad \z\gg1\quad\hbox{and}\quad \eta=\frac{kT}{72\pi\,c\,\d^2}\z^4,\quad \z\ll1,
\een
which are negligible quantities for a suitable $\z$.

Ultimately we analyze the shear viscosity $\mu$ defined in Eq. (\ref{17}b). From \cite{KM} we have that
\ben\label{22}
\mu=\frac{\mathcal{K}_{11}-2\mathcal{K}_{12}+\mathcal{K}_{22}}{\mathcal{K}_{11}\mathcal{K}_{22}-\mathcal{K}_{12}^2}.
\een
In figure 3 we see the graphics of the shear viscosity for a single gas in black (when $m_1=m_2=m, \n_1=\n_2=\n, \sigma_{11}=\sigma_{22}=\sigma_{12}=\sigma$) and in colored curves we set $\epsilon=0.01$ and $\xi=0.1$ and different concentrations.
\begin{figure}[h]
  \centerline{\includegraphics[width=270pt]{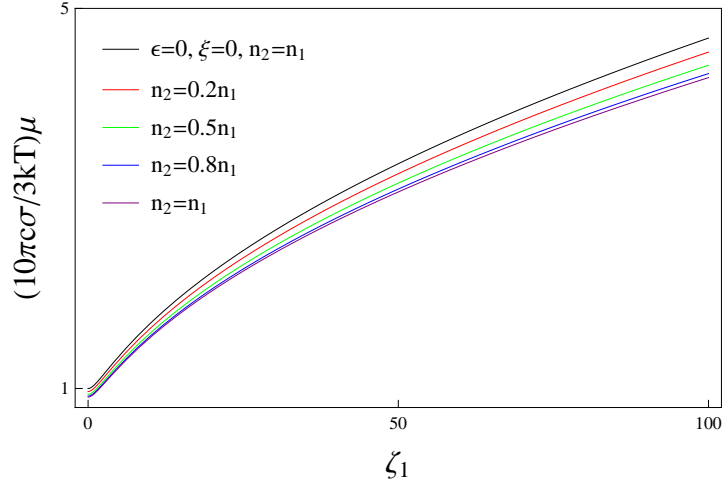}}
  \caption{(color on line) Shear viscosity (dimensionless) as a function of  $\z_1=m_1 c^2/kT$. Black line represents the single fluid whereas colored curves are plotted with $\epsilon=0.01$ and $\xi=0.1$ and different concentrations.}
\end{figure}
We can infer from figure 3 that the dimensionless shear viscosity coefficient assume large values in the non-relativistic limiting case $\z\gg1$ than those corresponding to the ultra-relativistic limiting case $\z\ll1$. We can conclude that the $\mu$ decreases by increasing the concentration of the constituent labeled by 2 with respect to the one labeled by 1 as it happens with the other coefficients.
For the case os a single component we consider that the rest masses, the particle number densities and  the differential cross-sections of both constituents are the same, then we get that the shear viscosity reduces to its known expression \cite{CK,GLW},
\ben\label{23}
\mu=\frac{15}{64\pi}\frac{kT}{c\sigma}\frac{\z^4K_3(\z)^2}{(2+15\z^2)K_2(2\z)+(3\z^3+49\z)K_3(2\z)}.
\een
The cases of low (non-relativistic $\z\gg1$) and high temperatures (ultra-relativistic $\z\ll1$) are obtained as
\ben\label{24}
\mu=\frac5{16\d^2}\sqrt{\frac{mkT}\pi} \left(1+\frac{25}{16\z}+\dots\right), \qquad \z\gg1,
\qquad
\mu= \frac{6kT}{5\pi\,c\,\d^2}\left(1+\frac{\z^2}{20}+\dots\right),\qquad\z\ll1.
\een

\section{FINAL REMARKS}

In this work we have shown analytical expressions for the transport coefficients of thermal conductivity, bulk and shear viscosities. We showed with the help of graphics that these coefficients decrease respect to the single component one when the concentration of the species 2 is less that the one of species 1. This last happens when the rest mass of the particle of species 2 is a little bigger than the rest mass of the particles of species 1, $m_2=m_1(1+\epsilon)$.

A very interesting issue to underline, as has been shown in the literature  \cite{KS,KD}, is the dependence of the transport coefficients on the gravitational potential. To do so, we can write the last two terms of Eq. (\ref{6.5}) by taking the sum over species and substituting the constitutive equations for the dynamical pressure and pressure deviator tensor:
\ben
\mathcal{P}^{\mu\nu}=-\p\Delta^{\mu\nu}+\eta \nabla_\gamma U^\gamma \Delta^{\mu\nu}+2\mu \left( \frac{\Delta^\mu_\sigma\Delta^\nu_\tau+\Delta^\nu_\sigma\Delta^\mu_\tau}2-\frac{\Delta^{\mu\nu}\Delta_{\sigma\tau}}3\right)\partial^\sigma U^\tau.
\een
Then, in Cartesian coordinates by using Eq. (\ref{11}b) we have
\ben\label{28}
\mathcal{P}^{ij}=\left[\widetilde\p
-\widetilde\eta\frac{\partial U^k}{\partial x^k}\right]\delta^{ij}-\widetilde\mu\left[\frac{\partial U^i}{\partial x_j}+\frac{\partial U^j}{\partial x_i}-\frac23\frac{\partial U^k}{\partial x^k}\delta^{ij}\right].
\een
Here we identify the quantities
\ben\label{29}
\widetilde\p=\frac\p{\left(1-\frac{\Phi}{2c^2}\right)^4},\qquad
\;\widetilde\eta=\frac{\eta}{\left(1-\frac{\Phi}{2c^2}\right)^8},\;\qquad
\widetilde\mu=\frac\mu{\left(1-\frac{\Phi}{2c^2}\right)^8}.
\een

On the other hand, the Fourier law in the same coordinate system reads
\ben\label{30}
\q^i=-\widetilde\lambda\frac{\partial \mathcal{T}}{\partial x_i},\qquad \hbox{with the thermal conductivity}\qquad \widetilde\lambda=\frac\lambda{\left(1-\frac{\Phi}{2c^2}\right)^4},
\een
and
\ben\lb{d3a}
\frac{\partial \mathcal{T}}{\partial x_i}&=&\frac{\partial T}{\partial x_i}-\frac{T}{c^2}\left(\dot U^i-\frac1{1-\Phi^2/4c^4}\frac{\partial \Phi}{\partial x_i}\right),
\een
where $\dot U^i$ denotes the acceleration.

Note that the quantity $\widetilde{\p}$ from equation (\ref{28}) is playing the role of the pressure and has a factor that depends on the gravitational potential. If we expand for $\Phi/c^2\ll1$ as we did for equation (\ref{12}) we recover the first post Newtonian approximation for the pressure as given by Weinberg \cite{Wein} or in the recent work  by using the Maxwell-J\"uttner distribution function in the 1PN formalism \cite{KRW}. So we have:

\ben\label{32}
\widetilde\p=\p\left[1+\frac{2\Phi}{c^2} +\mathcal{O}\left(\frac{\Phi^2}{c^4}\right)\right].
\een

Additionally, we can write the one-species and non-relativistic (low temperatures, $\z\gg1$) limits of Eqs. (\ref{29}b), (\ref{29}c) and (\ref{30}b) for a weak gravitational potential, yielding
\ben\label{32}
\widetilde\eta&=&\frac{25}{64\d^2\z^2}\sqrt{\frac{mkT}\pi}\left(1- \frac{183}{16}\frac{1}{\z^2}+\cdots\right) \left(1-\frac{4GM}{c^2 r}+\frac{9(GM)^2}{c^4 r^2}-\cdots\right)\\
\widetilde\mu&=&\frac5{16\d^2}\sqrt{\frac{mkT}\pi} \left(1+\frac{25}{16\z}+\dots\right)\left(1-\frac{4GM}{c^2 r}+\frac{9(GM)^2}{c^4 r^2}-\cdots\right)\\
\widetilde\lambda&=&\frac{75}{64\d^2}\frac{k}{m}\sqrt{\frac{mkT}\pi} \left(1+\frac{13}{16\z}+\dots\right)\left(1-\frac{2GM}{c^2 r}+\frac{5(GM)^2}{2c^4 r^2}-\cdots\right).
\een

We can conclude that this dependence on $\Phi=-GM/r$ decreases the value of the transport coefficients, but for non compact objects it can be a negligible contribution. The value of such quantity $\vert\Phi(R)\vert/c^2$, being $R$ the radius of the massive object, is about $7 \times 10^{-10}$ for the Earth, $2.2 \times 10^{-6}$ for the Sun. Other situations present a more suitable value as $2.8 \times 10^{-4}$ for a white dwarf or $7.5 \times 10^{-2}$ for a neutron star.

\section{ACKNOWLEDGMENTS}
The research of VM was supported by the Consejo Nacional de Ciencia y Tecnolog\'ia (CONACyT), M\'exico and  GMK  by the Conselho Nacional de Desenvolvimento Cient\'ifico e Tecnol\'ogico (CNPq), Brazil.


\end{document}